\documentclass[3p,preprint,12pt]{elsarticle}

\usepackage{hyperref}
\usepackage{amsmath}
\usepackage{bm}
\usepackage{amsbsy}
\usepackage{amssymb}
\usepackage{pdflscape}
\usepackage{color,soul} 

\journal{arXiv}

\begin{document}

\begin{frontmatter}
	
\title{Conformation Controllable Inelastic Charge Transport and Shot Noise Behavior in Metal-String Single Molecular Devices}

\author[1,2]{Talem Rebeda Roy}

\author[1,2]{Arijit Sen\corref{cor1}%
}
\ead{arijits@srmist.edu.in}

\cortext[cor1]{Corresponding author}

\address[1]{SRM Research Institute, SRM Institute of Science and Technology, Chennai 603203, India}
\address[2]{Department of Physics and Nanotechnology, SRMIST, Chennai 603203, India}

\begin{abstract}
It is often intriguing experimentally to take stock of how conformational changes in the device configuration may impact the overall charge transport behavior of single-molecule junctions. Based on the allied approach of density functional theory and non-equilibrium Green's function formalism, we explore here the effect of junction heterogeneity on inelastic charge transport in various metal-string based single-molecule devices. The constituent active elements being sensitive to the resonant levels, transition metal centers are found to  influence stretching, bending, and torsional excitation modes, while rocking and scissoring modes are controlled largely by the axial ligands. For certain molecular conformations and electrode orientations, phonon-assisted quantum interference effect may crop up, leading to the suppression of higher wavenumber vibrational modes. The resulting inelastic spectra are likely to take the shape of dominant Fano resonance or anti-resonance, depending on whether phonons are emitted or absorbed. Such nanoscale quantum interference effect is manifested especially in those metal-string molecular junctions for which the  energy gap (between  localized and delocalized virtual states) lies well within the optical phonon energies ($\Delta{E}_{|HOMO-LUMO|} <$ 40 meV). It also turns out that single molecular shot noise can exhibit nearly Poissonian behavior if the inter-channel tunneling through frontier orbitals is accompanied by phonon absorption or emission following a slow relaxation process. Our results thus suggest that charge transport properties across metal-string complexes can be potentially tuned by selective architecture of the metal centers and also, by preferred orientation of nanoscale electrodes in a bid to build up molecular devices with desirable controllability.
\end{abstract}


\end{frontmatter}

\section{Introduction}

To realize the successful miniaturization of nanoelectronic devices, single-molecule junctions have widely been studied\cite{Kushmerick2004,Joshua2011,Jia2015,Nichols2015,Perrin2015,Bessis2016,Xiang2016,garner2018} over the years at various levels from molecular imaging to molecular manipulation. Quantum transport properties of these junctions help us in understanding their obscure behavior by shedding light on various structure-function relationships that are often central to molecular electronics. Hetero-junctions at nanoscale are in general constructed by anchoring the molecular moieties through commonly used linkers such as thiol, amine or carboxylic-acid groups\cite{Chen2006,Sen2010,Kaliginedi2014,Youngsang2016,Sen2013,Arroyo2013,Shao2016,ROY2018}. In recent years, a different type of junction conformation based on metal-string complexes has also come to the fore \cite{Niskanen2012,Lai2010,Huang2011,Mohan2012,Huang2014} having potential applications as nanoscale interconnects. The linkers in such systems are made up of ambidentate \textit{isothiocyanate} (-NCS) ligands. An important aspect of such organometallic complexes is that the bond lengths between adjacent metal centers are usually very short, leading to strong interactions in the moieties\cite{Niskanen2012}. Metal-string complexes can form stable molecular junctions which in turn facilitate the charge transport\cite{Huang2014,Hsiao2008,Lai2008,Chiu2014,lin2004,chenPeng2006,shih2010} across it. Conductance of these junctions depends on both the electronic delocalization and the coupling of metal centers along the single-stranded metal-string. Besides, the electron spin levels can be split even in the absence of an external magnetic field \cite{Huang2014}. Metal-string complexes can thus have the right ability to tune the electronic conductance which is often useful for device applications. 

Electrons in a usual two-probe setup have certain probability to tunnel through a molecular bridge either elastically or inelastically under the influence of a finite bias \cite{Kamil2007,Reed2008,Joshua2010,Okabayashi2010}. The inelastic carrier transmission occurs when an inelastic electron tunneling lane correlates with the phonon excitation. At low temperatures, thermal excitation of phonons remains almost negligible but when the applied bias reaches the threshold voltage, $V_{th}$=$\hbar\omega/e$, a phonon of energy $ \hbar\omega $ may emanate \textit{via} the electron-phonon interaction. Inelastic electron tunneling spectroscopy (IETS) thus helps to fingerprint the junction molecules by way of molecular vibrations which are generated out of these electron-phonon interactions \cite{Jacob2013,Gunst2016}. Besides, IETS signals are also sensitive to molecular conformations and contact geometry. Though several important studies on the elastic charge transport properties of metal-string molecular junctions have been reported\cite{Niskanen2012,Lai2010,Huang2011,Mohan2012,Huang2014,Hsiao2008,Lai2008,Chiu2014,ROY2019}, vibrationally induced quantum kinetics in these systems has not yet been fully explored. However, mode-selective phononic control of charge transport in such kind of metalloligand wires may herald remarkable opportunities in device engineering at nanoscale.

In this work, the effect of metal-string conformations on the phonon-assisted tunneling of electrons is investigated from \textit{first-principles} for various single-molecule junctions comprising metal centers surrounded by ligands as active elements. The bias dependence of elastic as well as inelastic shot noise in such heterojunctions is subsequently discussed to better understand the inter-channel resonant tunneling of electrons, especially in the presence of diverse electrode-molecule coupling strengths. Metal centers appear to considerably impact the inelastic tunneling of electrons and hence, the overall charge transport across various homo- and hetero-nuclear metal-string moieties.

\section{Model and method}

As chosen for the present study, all the metal-string complexes comprised homo- and hetero-nuclear metal centers of type $[M]_{3}(dpa)_{4}$(NCS)$_{2}$ and $[M-M-M^{\prime}](dpa)_{4}$(NCS)$_{2}$ respectively, where $M$ = \{Co, Cr\} and $M^{\prime}$ = Ru. Further, each metal-string was capped with $isothiocyanate$ in the form of axial ligands, at both ends which eventually bridged the  gold electrodes through $thiol$ linkers. Such robust yet tunable binding\cite{hakkinen2012gold} between the ligand and gold atoms in the electrodes is likely to play an important role in improving the charge transport behavior in these single-molecule junctions \cite{Jacob2013,Vitali2010}. Since each metal-string under consideration consists of only transition metal ions, the high density of states near the Fermi level enhances the carrier transport \cite{Sen2010,Pontes2008}. The conductance values are often seen to vary appreciably in such heterojunctions depending on the kind of metal centers and also, the type of bonding in the metal-string \cite{Hsiao2008,Lai2008,Chiu2014}. Two different forms of electrode orientations, ${viz.}$ Au(111) and Au(100) nanowires (NWs) were chosen, so as to have symmetric as well as asymmetric single-molecule junctions. In the symmetric electrode configurations, both the electrodes on either side of the molecular moiety consisted of Au(111)NWs only, while in the asymmetric case, one of the electrodes was allowed to be formed by Au(100)NWs. 

Various device configurations, as described in Figure 1a-c, were relaxed by keeping six layers of nanowires fixed on both sides of the junction while allowing only the region comprising gold adatoms along with the metal-string complex to move, until the residual forces became smaller than 0.02 eV/\AA{}. This way, the symmetric electrode based nanojunctions contained as many as 185 atoms while the asymmetric ones were composed of 167 atoms. Each optimization procedure was carried out using the density functional theory (DFT), based on localized orbitals, as implemented in the SIESTA package\cite{SIESTA2002}. Exchange and correlation effects were treated within the generalized gradient approximation (GGA) by adopting Perdew-Burke-Ernzerhof (PBE) functionals \cite{Perdew1996}. Brillouin zone  sampling of $1\times 1\times 150$ $k$-points was used during the device optimization with an energy cut-off of 200 Ry, while the basis set was treated using double-$\zeta$ with polarization (DZP). In the optimized geometry, the intra-string metal-metal (M-M) bond lengths and also, metal-ligand bond distances turn out to be in good accord with the available measured data\cite{Huang2011,Hsiao2008}, as evident from Table S1 (see the Supplementary). Once all the device geometries got optimized, \textit{first-principles}  quantum transport calculations were performed for each junction geometry by integrating DFT with the non-equilibrium Green's function (NEGF) formalism  to take into account the open boundary conditions in a possible realistic way\cite{Taylor2001,Brandbyge2002,Atk2015}. We allowed each device to possess a scattering region extended over six layers of nanowires (serving as buffer layers) for semi-infinite electrodes on both sides of the respective molecular moiety in a two-probe setup\cite{Sen2010,Sen2013,ROY2018}. The device Hamiltonian was then expanded \cite{Taylor2001,Brandbyge2002,Sergueev2005} in real space by way of non-orthogonal localized pseudoatomic orbitals with the DZP basis set for respective individual elements. During the self-consistency cycle, norm-conserving Troullier-Martins pseudopotentials\cite{TM} were used to describe the effect of ion cores on the valence electrons. Self-consistency was achieved by using 300 \textit{k}-points along the transport direction.  We made use of unperturbed retarded (advanced) Green's functions $\textnormal{\bf{G}}^{r}(\textnormal{\bf{G}}^{a}=\textnormal{\bf{G}}^{r\dagger})$ construed in the scattering region, as depicted in Figure 1a-b. 

From the standard Landauer-B\"{u}ttiker formula, the elastic transmission matrix for a molecular junction is given by\cite{Taylor2001,Brandbyge2002} 
\begin{equation}
\textnormal{\bf{T}} = \bm{\Gamma} _{L}\textnormal{\bf{G}}^r\boldsymbol{\Gamma}_{R}{\textnormal{\bf{G}}^a},
\end{equation} 
with \textbf{$\boldsymbol{\Gamma}_{\textnormal L(R)}$} being the level broadening due to the respective electrode \textit{L(R)}. One conventional yet effective way to numerically estimate the inelastic current-voltage characteristics, electron-phonon couplings as well as vibrational frequencies is based on the so-called self-consistent Born approximation (SCBA). But a major difficulty in solving SCBA within the framework of the density functional theory (DFT) for as large and complex systems as what has been reported in the present work rests mainly with vast computational costs. However, the computational burden can be reduced to a good extent by simplifying calculations with the lowest order expansion (LOE) of the SCBA formalism\cite{Thomas2007}. Further improvements can be made by adopting an extended lowest order expansion (XLOE) method\cite{Gunst2016}, in which the Green’s functions as well as  the self-energies for the transmission function are evaluated at $\epsilon_F \pm \hbar\omega$. In XLOE, the equations are expanded to the lowest order in the electron-phonon self-energies, $\sum_\lambda$ and to the second order in electron-phonon coupling matrices, $\bf{M}_\lambda$ for the device region to predict the IETS signals which are prominent only when they are close to the excitation threshold. The Kohn-Sham Hamiltonian matrices, $\textnormal{\bf{H}(\bf{Q})}\equiv \big\{\big\{\big\langle i|\hat{H}_e|j\big\rangle\big\}\big\}$, for each ionic displacement are used to evaluate the \textit{e-ph} coupling matrices $\bf{M}_{\lambda}\equiv \big\{\big\{M_{ij}^\lambda\big\}\big\}$, with $M^\lambda_{ij}=\sum_{Iv}\bigg\langle i\bigg|\frac{\partial\hat{\bf{H}}_e}{\partial {\bf{Q}}_{Iv}} \bigg|j \bigg\rangle_{Q=0}v^\lambda_{Iv}\sqrt{\frac{\hbar}{2{\bf{M}}_I\omega_\lambda}}$\cite{Thomas2007}. Here, ${\bf{Q}}_{Iv}$ refer to the nuclear coordinates of the $I^{th}$ ion and ${\bf{M}}_I$, its mass while $v^\lambda_{Iv}$ denotes the ionic displacement.

The electron-phonon coupling in single molecular junctions are usually weaker as the electron hopping renders negligible contributions towards the charge transport so that the through-bond tunneling  remains prominent. Calculated IETS for the standard OPE based molecular junctions\cite{Thomas2007} and also, for BDT as well as  \textit{n}ADT molecular junctions\cite{Jing2014}, as obtained by exploiting the DFT-NEGF-LOE formalism, have been observed to be in good conformity with the experimental data. Hence, we have adopted this method for the present set of metal-complex molecular junctions. However, the extended LOE is considered here as it allows for the energy variation in the electronic structure on the scale of phonon energy. Inelastic correction to the tunneling current was therefore extracted from each vibrational mode by way of extended \textit{lowest order expansion} method\cite{Bessis2016,Gunst2016,Jing2014,Foti2015,Rasmus2015} to mainly take care of rapid vibrations near the electronic resonance in the weak electron-phonon (\textit{e-ph}) interaction limit. The tunneling current on being expanded to the second order in the \textit{e-ph} coupling matrix, $\bf{M}_{\lambda}$, can be expressed as the sum of elastic ($el$) and inelastic ($inel$) components such  that\cite{Bessis2016,Jing2014}
\begin{equation}
I(V,T) = I_{el}(V,T) + \sum_{\lambda}I_{inel}(V,\omega_{\lambda},T),
\end{equation} 
where $I_{el}(V,T)$ typifies the pure elastic current, as obtained from the Landauer-B\"{u}ttiker formula, while  $\sum_{\lambda}I_{inel}(V,\omega{_\lambda},T)$ refers to the inelastic contribution coming from each phonon mode, as indexed by $\lambda$,  with energy $\hbar\omega_{\lambda}$. 

Since the extended LOE takes cognizance of the energy variation in the electronic structure on the  scale of phonon energy, both the current and its second derivative with respect to the bias ($V$) can be written as the sum of two analytical functions in the following form\cite{Foti2015}
\begin{equation}
I^{LOE}(V,T) = \sum_{\lambda} I^{sym}_{\lambda}(V,\omega_{\lambda},T)T^{sym}_{\lambda}(\epsilon) + I^{asym}_{\lambda}(V,\omega_{\lambda},T)T^{asym}_{\lambda}(\epsilon), 
\end{equation} 
where \textit{symmetric} as well as \textit{asymmetric} contributions to the inelastic current in atomic units  (\textit{e = $\hbar$} = 1) can be approximated as\cite{Gunst2016,Jing2014,Foti2015,Rasmus2015}
\begin{equation} 
I^{sym}_{\lambda}(V,\omega_{\lambda},T) \equiv \frac{G_{0}}{2}\sum_{\sigma = \pm 1}\sigma(\omega_{\lambda} + \sigma V)\Big( \coth \frac{\omega_{\lambda}}{2k_{B}T} - \coth \frac{\omega_{\lambda}+\sigma V}{2k_{B}T}\Big)
\end{equation} 
and 
\begin{equation} 
\begin{aligned}
I^{asym}_{\lambda}(V,\omega_{\lambda},T) \equiv {} \frac{G_{0}}{2}\int_{-\infty}^{+\infty} d\epsilon \mathcal{H}\{f(\epsilon - \omega_{\lambda} ) - f(\epsilon + \omega_{\lambda}) \}(\epsilon)\\
[f(\epsilon - eV) - f(\epsilon)], 
\end{aligned}
\end{equation}
where $\mathcal{H}$ denotes the Hilbert transform, $f$(...) the Fermi function, $G_{0}$ (=2$e^{2}/\hbar$) the conductance quantum. While $I^{sym}_{\lambda}(V,\omega_{\lambda},T)$ yields symmetric conductance steps at vibrational energies, $I^{asym}_{\lambda}(V,\omega_{\lambda},T)$ results in asymmetric peaks/dips in the conductance with respect to the bias inversion. The terms associated with $\coth$ in Eq. (4) often lead to sharp peaks in the inelastic electron tunneling spectra around $|V| = \omega_{\lambda}$, having broadening of the order of $k_{B}T$. On the other hand, the IETS signal amplitudes, $\alpha_{\lambda}$ and $\beta_{\lambda}$, essentially represent the electron-phonon coupling, as given by\cite{Gunst2016,Jing2014,Foti2015,Rasmus2015}
\begin{equation}
T^{sym}_{\lambda}(\epsilon) = \textnormal{Tr} [\textnormal{\bf{M}}_{\lambda}\tilde{\textnormal{\bf{A}}}_{L}(\mu_{L})\textnormal{\bf{M}}_{\lambda}\textnormal{\bf{A}}_{R}(\mu_{R})] + \Im B_{\lambda},
\end{equation}
and 
\begin{equation}
T^{asym}_{\lambda}(\epsilon) = 2 \Re B_{\lambda},
\end{equation}
with $\tilde{\textnormal{\bf{A}}}_{\alpha}(\epsilon) = \textnormal{\bf{G}}^{a}(\epsilon) \bm{\Gamma}_{\alpha}(\epsilon)\textnormal{\bf{G}}^{r}(\epsilon)$ being the time-reversed form of the spectral density matrix, $\textnormal{\bf{A}}_{\alpha}(\epsilon)$, for the propagating states, and $\mu_{L(R)}$, the chemical potential of the left(right) electrodes such that $\mu_{R} = \mu_{L} \pm \hbar \omega_{\lambda}$. Further, $\Im$ and $\Re$ denote respectively the \textit{imaginary} and \textit{real} parts associated with the interference term $B_{\lambda}$, defined as\cite{Gunst2016,Jing2014,Foti2015,Rasmus2015}  
\begin{equation}
\begin{aligned}
B_{\lambda} = {} & \textnormal{Tr} [\textnormal{\bf{M}}_{\lambda}\textnormal{\bf{A}}_{R}(\mu_{L})\bm\Gamma_{L}(\mu_{L})\textnormal{\bf{G}}^{r}(\mu_{L})\textnormal{\bf{M}}_{\lambda}\textnormal{\bf{A}}_{R}(\mu_{R}) - \textnormal{\bf{M}}_{\lambda}\textnormal{\bf{G}}^{a}(\mu_{R})\bm\Gamma_{L}(\mu_{R}) \\
& \times \textnormal{\bf{A}}_{R}(\mu_{R})\textnormal{\bf{M}}_{\lambda}\textnormal{\bf{A}}_{L}(\mu_{L})],
\end{aligned}
\end{equation}

The IETS amplitude is expounded as the ratio of second and first derivative in the tunneling current ($I$) with respect to the bias ($V$) such that\cite{Foti2015,Rasmus2015} 
\begin{equation}
\textnormal{IETS} = \frac{d^{2}I^{LOE}(V,T)/dV^{2}}{dI^{LOE}(V,T)/dV}   
\end{equation}
By assuming the \textit{e-ph} coupling as perturbation on the tunneling current, IETS are finally evaluated for respective molecular junctions within the framework of extended DFT-NEGF-LOE formalism\cite{Gunst2016}. 

\section{Results and discussion}
\subsection{Inelastic electron tunneling}

In regard to a junction setup especially at nanoscale, inelastic electron tunneling spectroscopy serves as a cardinal approach for understanding the nature and effect of molecular motion during the process of electron tunneling. In the present study, a set of single-molecule junctions (SMJs) comprising both homo- and hetero-nuclear metal-string complexes are chosen, which bridge the Au(111)NWs initially in a symmetric way, as depicted in Figure 1a. Moreover, for the inelastic charge transport study of asymmetrically coupled molecular junctions, we incorporate Au(100)NWs as one of the defining electrodes (see Figure 1b). As Figure 1c suggests, the homo-nuclear part has a linear assembly of metal centers in the form of either \textit{tri-chromium}: [Cr]$_{3}(dpa)_{4}$(NCS)$_{2}$ or \textit{tri-cobalt}: [Co]$_{3}(dpa)_{4}$(NCS)$_{2}$ complexes. The terminally substituted ruthenium ions in those lead to hetero-nuclear complexes such as [Cr-Cr-Ru]$(dpa)_{4}$(NCS)$_{2}$ or [Co-Co-Ru]$(dpa)_{4}$(NCS)$_{2}$. In contrast to the tri-ruthenium junctions where the current flows primarily through the $\pi$-symmetry conduction channels, it is the $\sigma$ framework that governs the electronic tunneling in the case of tri-chromium as well as tri-cobalt metal-string junctions\cite{Mohan2012}. Figure 2 shows a comparative study of IETS for those junctions where the current flows through the $\sigma$ framework only. Systematic substitution of the Ru atom rather allows us to better understand the behavior of current flow due to the coupling of $\sigma$ and $\pi$ channels. These metal-string molecular moieties are shown schematically in Figure 1a-b along with respective vibrational boxes within the scattering regions where all the atomic vibrations take place at specific bias voltages, $V \geq \hbar\omega/e$. The prominent vibrational degrees of freedom in such type of heterojunctions with characteristic energies within 100 meV turn out to be $torsional$ ($\tau$), $stretching$ ($\nu$), $rocking$ ($\rho$), and scissoring or $bending$ ($\delta$) modes respectively, as sketched in Figure 1d and also, enlisted in Tables 1 and 2. 

The in-plane $bending$ and $stretching$ modes occur due primarily to the  metal-metal (M-M) linkages (intra-string) while the out-of-plane modes stem from the metal-nitrogen (M-N) bonds. On the other hand, vibrations due to $isothiocyanate$ linkers lead to the in-plane $scissoring$ and $rocking$ modes. The \textit{dpa} ligands, being linked to a metal-string through nitrogen bonds, try to pull and push the metal-string allowing it to vibrate in tandem with the ligands which eventually result in $torsional$ modes within the metal-string. We assign the vibrational modes associated with the \textit{dpa} ligands according to the Wilson-Varsanyi terminology (WVT)\cite{Wilson1934,Varsanyi1974,Varsanyi1969} for \textit{benzene} rings, since the former consist of \textit{pyridyl} rings as shown in Figure 1e. According to this scheme, $\omega_{y}$ and  $\omega_{z}$ modes arise from the in-plane translational as well as the out-of-plane rotational degrees of freedom with respect to $y$- and $z$-axis respectively. Also, the in-plane $6b$ $stretching$ modes along with the out-of-plane $16b$ ring modes exhibit a close resemblance with the vibrations of the $pyridyl$ rings in multi-nuclear metal-string complexes \cite{Varsanyi1974,Varsanyi1969}. These are some of the degenerate pairs in $benzene$ with  $e_{2g}$ and $e_{2u}$ symmetry while $\omega_{y}$ signifies an out-of-plane $six$-$fold$ rotation of the $benzene$ ring (along the $y$-axis) with respect to the ring plane \cite{Varsanyi1974}. The $torsional$ modes come into play at low frequencies at which $\omega_{y}$ and $\omega_{z}$ modes dominate the vibrations in the ligand in a synergistic fashion, as mentioned earlier. It may be noted that for $benzene$ rings, the $6b$ $stretching$ mode is available at the fundamental frequency of 521 $cm^{-1}$ ($\sim$ 64.4 mV), which is close to our observations for the \textit{dpa} ligands (see Table 1).

From our calculated inelastic electron tunneling spectra (IETS) for \textit{tri-chromium} single-molecule junctions, the Cr-N $stretching$ modes appear at 295, 356, 401 and 657 $cm^{-1}$ ($\sim$ 36.5, 44.1, 49.8 and 81.4 mV)  which are close to the respective experimental values of 300, 334, 398 and 648 $cm^{-1}$ ($\sim$ 37.2, 41.6, 49.4 and 80.5 mV), as obtained from the surface-enhanced Raman spectroscopy (SERS) study by Hsiao \textit{et al} \cite{Hsiao2008}. As Figure 2a suggests, the peaks with mode indices 78 and 85, corresponding to the out-of-plane $stretching$ of Cr-N bonds, emerge with higher intensities. The peak splitting, resulting in two  closely separated modes, happens at respectively 356 and 401 $cm^{-1}$ ($\sim$ 44.1 and 49.8 mV) due  to the lifting of degeneracy for the $16b$ mode, which is quite in congruence with Ref. 22.  Figure 2b shows that the $pyridyl$-$pyridyl$ out-of-plane $twisting$ in $tri$-$cobalt$ single-molecule junctions may be initiated at vibrational frequencies lower than 201 $cm^{-1}$ ($\sim$ 24.9 mV), which is in good agreement with the observations made by Lai \textit{et al} \cite{Lai2008}. Further, the presence of $16b$ and $6b$ modes in the given frequency range augurs well for both the spectra derived from theoretical as well as experimental data. The high intensity peak for the  mode index 38 at 94 $cm^{-1}$ ($\sim$ 11.7 mV) in Table 1 corresponds to the $bending$ of Co-N bonds along with the $isothiocyanate$ linkers, and is due to the combination of various vibrational modes. 

As it turns out, the vibrational motion in the $pyridyl$ rings leads to the twisting of metal-metal (intra-string) bonds giving rise to $torsional$ modes. The translation of $pyridyl$ rings along the $z$ direction ($\omega_{z}$) enhances the intensity of this particular mode since the vibration is longitudinal to the electron flow \cite{Audrey2013}. Besides, the out-of-plane vibrations at respectively 201, 318, and 588 $cm^{-1}$ ($\sim$ 24.9, 39.4, and 72.9 mV) render negligible intensities in comparison with those at 69 and 94 $cm^{-1}$ ($\sim$ 8.6 and 11.7 mV). In sharp contrast to what happens in  $tri$-$chromium$ junctions, we observe here one dip in the lineshape of IETS, as evident from Figure 2b at 69 $cm^{-1}$ ($\sim$ 8.6 mV) within the given bias window. According to Persson and Baratoff \cite{Baratoff1987}, such kind of dips may emerge whenever reduction in the elastic current surpasses the inelastic contribution. In the following, we suggest that it is associated with the phonon absorption\cite{Bayman1981,Paulsson2008,Shimazaki2008}. For the hetero-nuclear system like [Cr-Cr-Ru]$(dpa)_{4}$(NCS)$_{2}$, as shown in Figure 2c, the high resonant peak at 350 $cm^{-1}$ ($\sim$ 43.4 mV) originates from the $stretching$ mode of Cr-Ru along the direction of electron transmission. Additionally, the $in$-$plane$ vibrations associated with the mode index 74 at 391 $cm^{-1}$ ($\sim$ 48.5 mV) lead to high resonance. As demonstrated in Figure 2d, the high resonant peaks at 419 $cm^{-1}$ ($\sim$ 51.9 mV) for the [Co-Co-Ru]$(dpa)_{4}$(NCS)$_{2}$ system  may be attributed to the out-of-plane $stretching$ of the Co-N bond as well as $bending$ in the $pyridyl$ rings. An on-resonance dip is further observed at 465 $cm^{-1}$ ($\sim$ 57.7 mV) as a result of phonon absorption\cite{Jing2014}.

We analyze the nature of IETS signals on the basis of \textit{partial device density of states} (PDDOS) in the first place, since the orbitals contributing to the electronic transmission of various single-molecule junctions tend to behave often differently. As Figure 2e-h suggests, the metal-string complexes exhibit high \textit{density of states} near the Fermi level. Besides, the \textit{local density of states} (LDOS) for the \textit{tri-chromium} complexes indicates the formation of $\pi$ bonds by the $3d_{zx}$ orbitals of the first and second metal centers, whereas $\delta$ bonds are formed by the $3d_{xy}$ orbital of the third Cr center. Further, the Cr-Cr-Cr string becomes more symmetric at ambient temperature \cite{Mohan2012,Chiu2014}. For the $tri$-$cobalt$ system, the prominent IETS peak at 94 $cm^{-1}$ ($\sim$ 11.70 mV) with mode index 38 may be attributed to the Co-$3d_{xy}$ orbital. Here, the $\sigma$ bonds, formed by the $3d_{z^2}$ orbital of the first Co center, dominate the transmission. On the other hand, for the second and third metal centers, the $\delta$ and $\pi$ bonds contribute largely to the electronic transmission. In both $tri$-$chromium$ or $tri$-$cobalt$ systems, the formation of localized $\pi$ ($d_{zx}$ and $d_{yz}$) or $\delta$ ($d_{xy}$) states help retain the linear bonding \cite{Mohan2012} of the respective metal core. However, in the case of hetero-nuclear metal-string junctions, the resonant peaks at 43.4 and 48.5 mV have the dominant orbital characters of $3d_{xy}$ ($3d_{zx}$) and $4d_{zx}$, stemming respectively from  Cr (Co) and Ru metal centers associated with [Cr-Cr-Ru]$(dpa)_{4}$(NCS)$_{2}$ ([Co-Co-Ru]$(dpa)_{4}$(NCS)$_{2}$) moieties. 

To investigate the IETS in asymmetrically coupled junctions, we resort to a different set of junction configurations in which Cr- as well as Co-based metal-string complexes are coupled to Au(100)NW electrodes on one end while to Au(111)NW electrodes on the other one. These configurations are chosen so as to understand the behavior of inelastic electron tunneling in multi-nuclear metal-string molecular junctions in the presence of asymmetric electrodes, and to this end, we begin as before with the homo-nuclear metal-strings. According to Figure 2i, the IETS signal of the $tri$-$chromium$ system, displays only one dip at 201 $cm^{-1}$ ($\sim$ 24.91 mV) though it vanishes with symmetric electrodes due likely to the absence of phonon absorption. In contrast, the IETS signal of the $tri$-$cobalt$ system (see Figure 2)) displays two dips at respectively 65 and 208 $cm^{-1}$ ($\sim$ 8.1 and 33.2 mV) in comparison with a single dip at 69$cm^{-1}$ ($\sim$ 8.6 mV) for the symmetric electrode configuration (see Figure 2b). The orbital contribution from Co$-3d_z^2$ (see Figure 2f), which is responsible for the strong IETS peak near the Fermi level in symmetrically coupled junctions, disappears in the density of states (DOS) of the $tri$-$cobalt$ system, once asymmetry is introduced in the electrodes (see Figure 2n). However, the situation reverses with the $tri$-$chromium$ system, where the metal-string DOS gets rather enhanced under the asymmetric electrode coupling (see Figure 2m). Once the crystallographic orientation of one of the two electrodes is changed, the effective Fermi level of the device shifts away from its earlier position, leading to certain changes in the device density of states (DDOS), as demonstrated in Figure 2. As a result, the orbitals contributing to the electronic transmission of various single-molecule junctions tend to behave quite differently. A pictorial table of these vibrational modes associated with symmetrical and asymmetrical junctions are given in Tables S2 and S3 respectively. 

The exclusion of dips in the IETS signal of [Cr-Cr-Ru]$(dpa)_{4}$(NCS)$_{2}$ system, as shown in Figure 2k indicates merely the absence of phonon absorption. The resonant intensities are, however, very low due to dominance of the $out$-$of$-$plane$ vibrations from the ligand. As portrayed in Figure 2o-p, ruthenium substitution considerably modifies the metal-string DOS in the asymmetrically coupled junctions though the impact is more with the Cr-based trinuclear systems than with the Co-based ones. For [Cr-Cr-Ru]$(dpa)_{4}$(NCS)$_{2}$ moiety, $DOS$ appears in the \textit{highest occupied molecular orbital} (HOMO) region at around -79.0 meV (not shown here) due to Cr$-3d_{zx}$ and Ru$-4d_{zx}$ metal-string states once coupled asymmetrically to an Au(100)NW electrode on one end. It becomes suppressed only when it gets coupled symmetrically on both sides to Au(111)NW electrodes. However, an opposite trend is observed for [Co-Co-Ru]$(dpa)_{4}$(NCS)$_{2}$ molecular moiety, where a large $DOS$ appears in the \textit{lowest unoccupied molecular orbital} (LUMO) region due to Co$-3d_{zx}$ and Ru$-4d_{zx}$ metal-string states, with asymmetrically coupled electrodes. IETS and PDDOS for [Cr-Ru-Cr]$(dpa)_{4}$(NCS)$_{2}$, [Cr-Ru-Ru]$(dpa)_{4}$(NCS)$_{2}$ and [Ru-Ru-Ru]$(dpa)_{4}$(NCS)$_{2}$ junctions are also shown in Figure S1 of supplementary information. The systematic replacement of Cr with Ru in the hetero-nuclear metal-string complexes usually increases the density of $d$-electrons in the transmission channels. However, for the \textit{tri-ruthenium} system, the presence of $sp^2$ hybridization between $\sigma$ state ($d_{z^2}$) of the first and $\pi$ state ($d_{xz}$) of the second Ru metal center eventually leads to the drop-off in $d$-electron contributions to the PDOS\cite{Mohan2012,Chiu2014,Niskanen2012}.  

Peaks and dips in the inelastic electron tunneling spectra (IETS) are often closely related to the chemical structure of molecules as well as electrodes due to diverse nature of the metal-molecule interaction, not only at the junction interface but also within the complex molecular moiety having metal centers. As the Table S4 suggests, for homo-nuclear tri-chromium metal-string complexes, the prominent vibrational modes are $\nu$ (out-of-plane $stretching$ of Cr-N) as well as $\delta$ (out-of-plane $bending$ of Cr-N) with the symmetric electrodes while only $\delta$ (out-of-plane $bending$ of Cr-N) with the asymmetric ones. Conversely, for homo-nuclear tri-cobalt metal-string complexes, the prominent vibrational modes appear to be $\nu$ (out-of-plane $stretching$ of Co-N) as well as $\delta$ (out-of-plane $bending$ of Co-N) with the asymmetric electrodes while only $\delta$ (out-of-plane $bending$ of Co-N) with the symmetric ones. On the other hand, for hetero-nuclear Cr-Cr-Ru metal-string complexes, the dominant modes are $\nu$ (out-of-plane $stretching$ of Cr-N) as well as $\delta$ (in-plane $bending$ of Cr-N) with the asymmetric electrodes while only $\delta$ (out-of-plane $bending$ of Cr-N) with the symmetric ones. Likewise, for hetero-nuclear Co-Co-Ru metal-string complexes, $\nu$ (out-of-plane $stretching$ of Co-N) as well as $\delta$ (in-plane $bending$ of Co-N) modes are prominent with the asymmetric electrodes while only $\nu$ (out-of-plane $stretching$ of Co-N) mode with the symmetric ones (see Table S4). From the Table S5, we further come across that the out-of-plane $stretching$ mode, $\nu$(M-N), is active only for the homo-nuclear metal-string junctions, while the in-plane $stretching$ modes, $\nu$’(M-M) and $\nu$”(M-M-M), dominate only for the hetero-nuclear systems. However, the out-of-plane $bending$ mode, $\delta$(M-N), turns out to be conspicuous for homo- as well as hetero-nuclear metal-string junctions, though the in-plane $bending$ mode, $\delta$' (M-M-M), dominates only for the homo-nuclear systems. 

Figure 3a-h display the transmission eigenstates at the Fermi level for respective metal-centers belonging to both homo- and hetero-nuclear junctions that bridge the symmetric as well as asymmetric nano-electrodes while the adjacent figures portray the schematics of the most prominent vibrational modes only. The atomic movements associated with $\omega_{z}$, $\omega_{y}$, $16b$ and $6b$ ring modes, which essentially capture the vibrational pattern in the $pyridyl$ rings, have already been outlined in Figure 1e, as per the WVT scheme. It turns out that the prominent peaks in the symmetrically coupled junctions stem from $\omega_{y}$, $\omega_{z}$, $\delta$, $\nu$ and $16b$ modes, while those in the asymmetrically coupled junctions arise out of $\omega_{y}$, $\omega_{z}$ and $\delta$ modes only.

\subsection{Fano resonance}

As we know, peaks and dips appear usually in the second derivative of the respective current-voltage (\textit{I-V}) characteristic curves whenever the electronic energy, inherent in the bias \textit{V}, harmonizes\cite{Galp2004} with the vibrational one. Nevertheless, some satellite resonant peaks may as well crop up even in the first derivative of \textit{I-V} plots for certain systems, espousing the onset of phonon-assisted charge transport (see Figure 4). In the present work, certain vibrational modes at respectively 295, 356, 402 and 457 $cm^{-1}$ ($\sim$ 36.6, 44.1, 49.8 and 56.7 mv), which are active (see Figure 2a) during the inelastic charge transport in SMJs comprising [Cr]$_{3}(dpa)_{4}$(NCS)$_{2}$ metal-string moieties interfaced with symmetric electrodes, get suppressed once the coupling turns asymmetric, as suggested by Figure 2i and 4e. However, such suppression of active vibrational modes at 268, 316 and 359 $cm^{-1}$ ($\sim$ 33.2, 39.4 and 44.5 mv) happens for [Co]$_{3}(dpa)_{4}$(NCS)$_{2}$ metal-string complexes if interfaced with rather symmetrically coupled electrodes, as demonstrated by Figures 2b and 4b. On the other hand, hetero-nuclear complexes such as [Co-Co-Ru]$(dpa)_{4}$(NCS)$_{2}$ go through similar kind of mode-suppression phenomena even with the symmetric electrode configuration, as apparent from Figure 2d and 4d. Although the IETS signal for molecular vibrations that are not directly involved in the electron-transport pathways can undergo mode suppression, such phenomena are also likely to occur as a result of quantum interference effects. According to a recent study made by Lykkebo \textit{et al}\cite{Lykkebo2014}, the non-overlapping of transmission channels stemming from quantum interference within the molecular motif may cause suppression of the corresponding vibrational modes. Metal centers in single-molecule junctions may thus have diverse yet discernible impact on the inelastic transmission whose resonant peaks or dips are often determined by the relative values of elastic and inelastic contributions to the net current near the phonon excitation threshold.

Figure 5 depicts a field plot on how the energy gap between HOMO and LUMO may be distributed as a function of the phonon wavenumber among various metal-string molecular junctions. The region within 295 $cm^{-1}$ ($\sim$ 36.6 mV) is found to be dominated by the \textit{stretching}, \textit{bending}, and \textit{torsional} modes respectively, while  the \textit{out-of-plane} \textit{16b} modes are dominant in the range of 296 - 550 $cm^{-1}$ ($\sim$ 36.7 - 68.1 mV). A third region between 551 and 800 $cm^{-1}$ ($\sim$ 68.2 and 99.3 mV) is  dominated exclusively by the \textit{in-plane} \textit{6b} modes. As it turns out, single-molecule metal-string junctions, prone to the phonon-assisted QI effect\cite{Bessis2016,anders2017,Salhani2017,Hartle2009,Hartle2011,Hartle2013}, could lie only in the redlined region of Figure 5. Here, the respective energy gaps of the two consecutive interfering states are very narrow (\textit{i.e.} $\bigtriangleup E_{|{HOMO} - {LUMO}|}$ $<$ 40 meV), lying well within the typical range of thermal phonons. However, the presence of energy gap between the two frontier orbitals within the given energy window is not sufficient for the phonon-induced QI to happen, since it is also necessary to have one of the orbitals in the localized form while the other to be delocalized. For instance, the asymmetric [Co]$_{3}(dpa)_{4}$(NCS)$_{2}$ junction fails to demonstrate QI since the frontier orbitals are all delocalized. The field plot thus helps us to qualitatively understand the basis for phonon-mediated QI phenomena occurring predominantly in three systems out of several metal-string molecular junctions under study. Also, it demonstrates how phonons may assist the QI to control the overall charge transport process.

With the symmetric electrode conformation, sharp intensity enhancement in the inelastic electron tunneling spectra is observed for [Co]$_{3}(dpa)_{4}$(NCS)$_{2}$ and [Co-Co-Ru]$(dpa)_{4}$(NCS)$_{2}$ heterojunctions while in the case of asymmetric electrodes, it happens only to [Cr]$_{3}(dpa)_{4}$(NCS)$_{2}$ systems (see Figure 2). Interestingly, these single-molecule junctions, being sensitive to the tunnel energy ($\varepsilon$), tend to display the characteristic Fano-type asymmetric lineshapes ($\varXi$) in respective IETS, as expressed by\cite{Zacharia2001} 

\begin{equation}
\varXi(\tilde{\varepsilon}) \propto \frac{(\tilde{\varepsilon}+q)^2}{\tilde{\varepsilon}^2 +1} \hspace{0.3cm}\textnormal{for}\hspace{0.3cm}\tilde{\varepsilon} = \frac{(\varepsilon - \varepsilon_{0})}{\Gamma/2},    
\end{equation} 

where \textit{q} denotes the Fano asymmetry parameter that determines the shape of resonance line,  while $\tilde{\varepsilon}$ represents the dimensionless resonance detuning, $\varepsilon_{0}$ the resonance level, and $\Gamma$ the resonance width. At the device level, it may be attributed to the quantum interference of the inelastic channel with the elastic one in which the direct tunneling is coupled with either \textit{excitation} or \textit{de-excitation} of a local vibrational mode within the molecular moiety. These above three unique systems will henceforth be referred to as \textit{s}-CoCoCo, \textit{s}-CoCoRu and \textit{a}-CrCrCr respectively for the sake of brevity. A fit of Eq. (10) to the IETS of \textit{s}-CoCoCo and \textit{s}-CoCoRu respectively yields ${|q|}$ $\approx$ 17.6 and ${|q|}$ $\approx$ 0.05, while IETS of \textit{a}-CrCrCr renders ${|q|}$ $\approx$ 3.6. It implies that the electron-phonon coupling strength is much stronger in \textit{s}-CoCoCo than in \textit{a}-CrCrCr and \textit{s}-CoCoRu.

The onset of Fano resonance \cite{Zacharia2001,Hasdeo2014} may be understood, albeit qualitatively, by way of a simplistic model \cite{Sen2010,Papadopoulos2006,Nozaki_2013}, so that the respective molecular moiety may be represented by a closely coupled two-level system, stemming from the two renormalized frontier orbitals (\textit{viz.} HOMO and LUMO). One of these states, being delocalized in nature, gets coupled with the left (right) electrodes \textit{via} the  coupling constant of $\gamma_{1}$($\gamma_{2}$), providing thus a direct channel for the incoming electrons. However, the inter-level coupling strength, as given by $t_{c}$, is mediated by phonons. On the other hand, the localized state, as assisted by phonons through either \textit{emission} or \textit{absorption}, gives rise to an  indirect pathway. Once these two transport pathways interfere at the band continuum provided by the semi-infinite quasi-1D electrodes, it results in the resonant suppression of the electronic transmission leading to the destructive QI, while the peaks may occur due to the constructive QI at their respective phonon excitation thresholds\cite{Lykkebo2014,Butzin2011,Ballmann2012,Li2016,Lambert2015,borges2016}. In the HOMO dominated transport junctions such as \textit{a}-CrCrCr and \textit{s}-CoCoRu (see Figures 6a and 6c), an incoming electron from the left electrode tunnels inelastically through a delocalized state near the renormalized HOMO by absorbing a phonon of certain energy ($\Omega_{1}$ or $\Omega_{3}$)  to reach the right electrode. This direct channel, responsible for the Breit-Wigner resonance, subsequently interferes at the band continuum with the indirect channel as provided by a localized state near the renormalized LUMO to result in the Fano resonance. However, in the  LUMO dominated transport junction of \textit{s}-CoCoCo, the inelastic tunneling occurs (see Figure 6b) through a delocalized state (rendering the direct channel) by emitting a phonon of energy $\Omega_{2}$ to reach the right electrode via a localized state (rendering the indirect channel) near the renormalized HOMO to reach the right electrode yielding in effect a Breit-Wigner-Fano (BWF) kind of a lineshape. In the wide-band limit (WBL) the model transmission may take the following form \cite{Papadopoulos2006,Nozaki_2013}:

\begin{equation}
\tau(\varepsilon) = \frac{4\gamma_{1}\gamma_{2}}{(\varepsilon-\varepsilon_{1} - \frac{t_c^2}{\varepsilon - \varepsilon_2})^2 + 4\bar{\gamma}^2}
\end{equation}
where $\bar{\gamma}=(\gamma_1+\gamma_2)/2$. Figures 6a-c indicate a reasonably good fit for the respective transmission profiles  with this model, where the two electronic levels, $\varepsilon_1$ and $\varepsilon_2$, serve as the leading transport channels. It may be noted that these two non-degenerate states as denoted by the vertical bars in the insets of transmission plots (see Figure 6) correspond neither to the peaks observed in the projected density of states nor to the renormalized HOMO/LUMO as denoted by the stars for isolated molecules. The latter have respective dominant orbital characters of $d_{xy}$ and $d_{yz}$ in \textit{a}-CrCrCr, $d_{zx}$ and $d_{xy}$ in \textit{s}-CoCoCo and $d_{zx}$ in the case of \textit{s}-CoCoRu system, as obtained by diagonalizing the device Hamiltonian matrix projected on the scattering region \cite{Sen2010}. From the respective model fit of Figures 6a-c, the coupling ratio ($t_{c}/\bar{\gamma}$) is estimated as 2.9 for \textit{a}-CrCrCr, 0.5 for \textit{s}-CoCoCo and 4.5 for \textit{s}-CoCoRu. A higher $t_{c}/\bar{\gamma}$, associated with \textit{s}-CoCoRu implies a stronger HOMO-LUMO coupling in this system than in \textit{a}-CrCrCr and \textit{s}-CoCoCo. However, higher value of $\bar{\gamma}$ accounts for stronger electrode-molecule coupling in \textit{a}-CrCrCr and \textit{s}-CoCoCo. A detailed analysis of LDOS at the device level as portrayed in the right panel of Figures 6 and S2 (renormalized LDOS) helps to shed light on why the phonon-assisted Fano resistance may occur in three out of eight heterojunctions under study, in tune with the field plot of Figure 5. Here, $d_{xy}$, $d_{yz}$ and $d_{zx}$ are the dominant orbital characters impacting each of the two levels, $\varepsilon_1$ and $\varepsilon_2$. The interaction of conducting electrons with the localized vibrational degrees of freedom can thus play an important role in the transport properties of single molecular junctions, which is in congruence with the previous reports\cite{Kushmerick2004,Youngsang2016,Troisi2006}.

\subsection{Shot noise characteristics}

A tunnel junction is often subjected to electronic shot noise stemming from the discretization of electrical charges due to direct tunneling events. In the coherent tunneling regime, it is thus important to understand the shot noise\cite{Kumar2012,Okazaki2013,vardimon2016,Tsutsui2010,Karimi2016} behavior at the molecular scale, especially when the conducting electrons interact with the vibrational degrees of freedom. The inter-channel interaction, coupled with phonon emission or absorption, may often inculcate the shot noise out of the current fluctuations\cite{Lambert2015} in single-molecule junctions. Additionally, whenever a new inelastic channel is opened up in the single-molecule junction, the local heating raises its temperature resulting in some thermal noise. The total current noise characteristics, $S_{I}(V)$, in absence of the \textit{e-ph} coupling may be calculated in terms of $n^{th}$ channel transmission probability ($\tau_{n}$) taken at the Fermi level for $N$ conducting channels such that\cite{Kumar2012,Okazaki2013,vardimon2016,Tsutsui2010,Karimi2016}
\begin{equation}
S_{I}(V)=4k_{B}TG_{0}\bigg[\sum_{n=1}^N\tau_{n}^{2} + X(V)\sum_{n=1}^N\tau_{n}(1-\tau_{n}) \bigg],
\end{equation}  
where $G_{0} = 2e^{2}/\hbar$ is the quantum of conductance, $V$ the applied bias and  $X(V) = (eV/2k_{B}T)\coth{(eV/2k_{B}T)}$ the control parameter. While the first term in Eq.(12) represents the Johnson-Nyquist thermal noise, the second one essentially accounts for the shot noise arising out of non-equilibrium distribution of electronic charges. 

Figure 7a-b shows how the finite-bias noise, $S_{I}(V)$, depends on the external bias in the absence of electron-phonon interactions. The linear dependence of shot noise on the applied bias is evident in both the cases. However, \textit{a}-CrCrCr experiences the total current noise, about two orders of magnitude, lower in value than what \textit{s}-CoCoCo does. A useful measure for the relative noise strength is often provided by the dimensionless ensemble-averaged Fano factor (\textit{F}) \cite{Okazaki2013,vardimon2016}, 
\begin{equation}
F=\frac{Y(V)}{X(V)-1},
\end{equation}
where $Y(V) = [S_{I}(V) - S_{I}(0)]/S_{I}(0)$, which represents the \textit{reduced excess noise} with  $S_{I}(0)$ being the thermal noise (at zero bias). From the linear fit of $Y(V)$ \textit{vs} $X(V)$ using Eq. (13), as illustrated in Figure 7c-d, the Fano factor, $F$ is estimated as 0.776 for \textit{s}-CoCoCo and 1.005 for \textit{a}-CrCrCr so that both these nanojunctions tend to exhibit nearly the full Poissonian noise characteristics (\textit{i.e.} $F \sim 1$). This implies overall the absence of electronic correlation between the tunneling events, although the electrons may encounter inelastic collisions. Poissonian behavior of the shot noise at the molecular scale may be construed within the Franck-Condon picture stemming from the resonant electronic transport between diverse charging states\cite{Koch2005,Sun2019,Tian2019}. Following the charge stability plots, as demonstrated in Figure 7(e-f), there is no overlap in the  Coulomb diamond edges around the zero bias. Such kind of conductance suppression at low bias happens mainly due to what is known as the Franck-Condon (FC) blockade, signifying the onset of electron-phonon coupling\cite{burzuri2014}. As we know, the FC blockade can not be lifted  by simply applying a gate voltage like what happens in case of the Coulomb blockade. However, the FC blockade may be overcome once the applied bias conforms to a certain threshold bias ($V_{FC}$) such that $V_{FC}\sim\lambda^2 \hbar\omega_0$, whence we can roughly estimate the electron-phonon coupling ($\lambda$) to be around 1.2 (i.e. $\lambda>1$) for both \textit{s}-CoCoCo and \textit{a}-CrCrCr systems.  

To assess further the phonon-assisted shot noise response\cite{Kumar2012,sumit}, we make use of its derivative with respect to the bias, $\delta \dot{S}(V)$ = $\partial_{V}(\delta {S}(V)$, so that the associated jump in the inelastic correction for the mode of $\lambda$ may take the following form\cite{avriller1}  
\begin{equation}
\Delta \dot{S_{\lambda}}/e \Delta G \approx \textnormal {Tr}\{(\textbf{1}-2\textbf{T})\textbf{T}^{LOE}_{\lambda}\},
\end{equation}
where $\Delta G = (2e^{2}/h) \sum_{\lambda} \textnormal{Tr} \{\textbf{T}^{LOE}_{\lambda}\}$ with \textbf{T}$^{LOE}_{\lambda}$ being the inelastic transmission matrix associated with $\lambda$, as obtained from Eqs. (6) and (7) within the lowest-order expansion. Figure 8a-h shows the plot of $\Delta \dot{S_{\lambda}}/e \Delta G$ as a function of the induced bias, corresponding to three prominent vibrational modes stemming from each of the eight molecular junctions under study. As it appears, jump in the inelastic correction to the shot-noise remains quite pronounced in hetero-nuclear junctions with asymmetric electrode conformations, due mainly to strong fluctuations in the locally excited vibrational modes. For the Cr-Cr-Ru metal-string being coupled to asymmetric electrodes, fluctuations in the local modes of $\omega_{y}$ and  $\omega_{z}$ at 104 $cm^{-1}$ ($\sim$ 12.9 mV) lead to an appreciably high jump in the inelastic correction to the shot-noise, see Figure 8g. However, for the Co-Co-Ru system being coupled to asymmetric electrodes, as depicted in Figure 8h, the respective threshold occurs due to equally localized character of $\omega_{y}$ and  $16b$, associated with all the principal modes at respectively 32, 180 and 334 $cm^{-1}$ ($\sim$ 3.9, 22.3, and 41.5 mV).

\begin{table*}[htb!]
\scriptsize
	\caption{Assignment of vibrational modes associated with the prominent IETS peaks for  homo- and hetero-nuclear metal-string molecular junctions coupled symmetrically to Au(111)NW  electrodes on both ends. The peak position of \textit{tri-chromium} and \textit{tri-cobalt} junctions are compared to various experimentally reported data\cite{Lai2010,Hsiao2008,Cortijo2018}. The suppression of vibrational modes in case of the \textit{s}-CoCoCo and \textit{s}-CoCoRu system is due mainly to the phonon-assisted quantum interference effect. Please refer to the article, as published in Applied Surface Science  (https://doi.org/10.1016/j.apsusc.2019.145196), for the details of this Table.}
\end{table*}

\begin{table*}[htb!]
\scriptsize
	\caption{Assignment of vibrational modes associated with the prominent IETS peaks for  homo- and hetero-nuclear metal-string molecular junctions coupled asymmetrically to Au(100)NW  electrode on one end while to Au(111)NW electrode on the other. The peak position of \textit{tri-chromium} and \textit{tri-cobalt} junctions are compared to various experimentally reported data\cite{Lai2010,Hsiao2008,Cortijo2018}. The suppression of vibrational modes in case of the \textit{a}-CrCrCr system is due mainly to the phonon-assisted quantum interference effect Please refer to the article, as published in Applied Surface Science  (https://doi.org/10.1016/j.apsusc.2019.145196), for the details of this Table.}
\end{table*}

\section{Conclusions}
We have analyzed here from \textit{first-principles} the effect of electrode as well as metal-string configuration on the phonon-assisted  tunneling of electrons  across various tri-nuclear metal-string molecular moieties. It has been observed that the stretching, bending,  and torsional vibrational modes are strongly affected by individual metal centers, while the axial ligands can mostly control the rocking and scissoring modes. Our results further suggest that metal centers in single-molecule junctions can have diverse yet discernible impact on the electronic transmission whose resonant peaks or dips are often determined by the relative values of elastic and inelastic contributions to the net current near the phonon excitation threshold. We attribute the Fano-type asymmetric line shapes in the inelastic electron tunneling spectra of \textit{s}-CoCoCo, \textit{s}-CoCoRu, \textit{a}-CrCrCr systems to the quantum interference effect of the indirect inelastic channel with the direct elastic one  within the energy range of optical phonons which eventually results in the suppression of higher wavenumber vibrational modes. Our charge transport analysis further implies that the single-molecule elastic shot noise can exhibit nearly Poissonian behavior if the inter-channel tunneling of electrons from HOMO(LUMO) to LUMO(HOMO) is accompanied by phonon absorption (emission) following a slow relaxation process. Phonon-assisted quantum interference effects at the molecular-scale, stemming from the interactions between incoming electronic states and internal phonon degrees of freedom, may also lead to a resonant enhancement in the vibrationally induced decoherence effects, which can be potentially harnessed for tailoring molecular nanodevices with wider functionalities.

\begin{figure}
	\centering
	\includegraphics[scale=0.81]{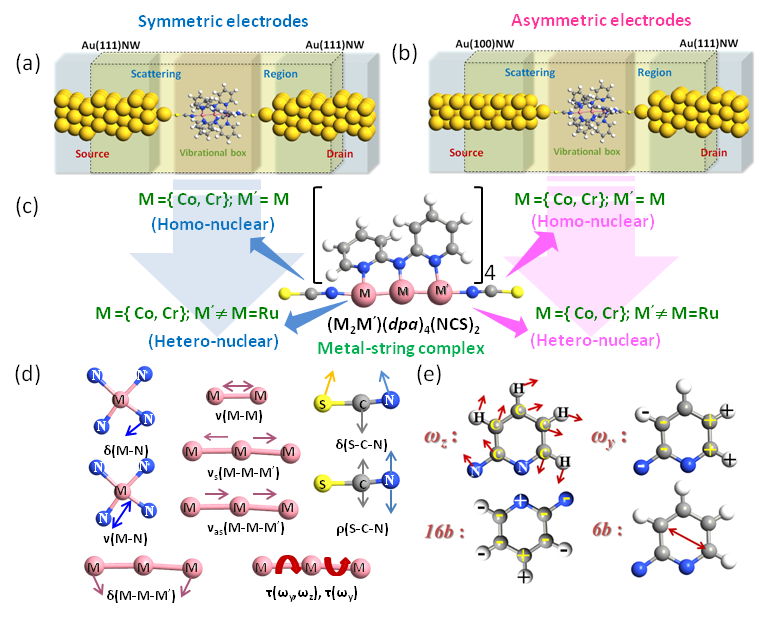}
	\caption{Please refer to https://doi.org/10.1016/j.apsusc.2019.145196 for further details.}    
\end{figure}

\begin{figure}
	\centering
	\includegraphics[scale=0.18]{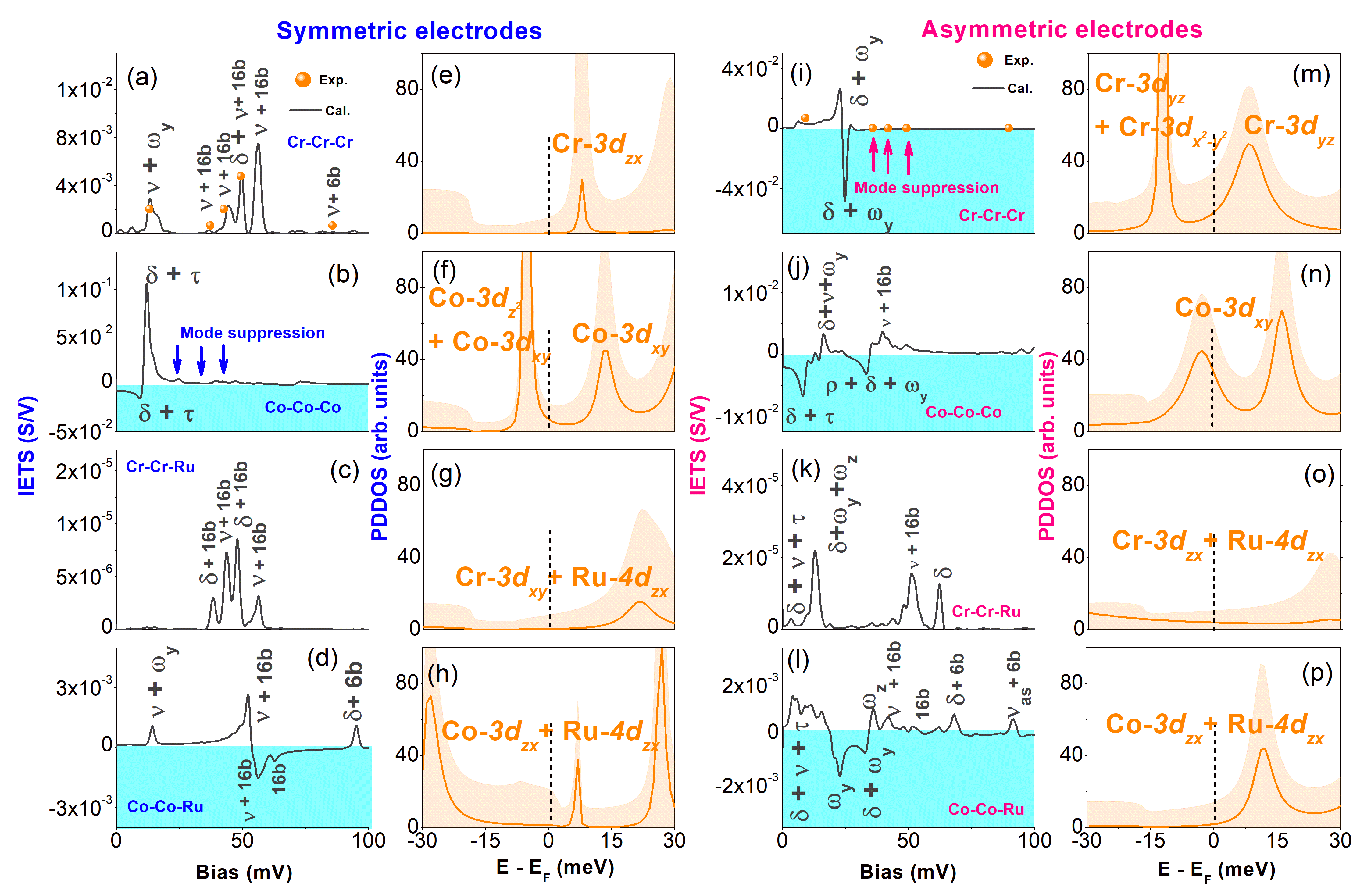}
	\caption{Please refer to https://doi.org/10.1016/j.apsusc.2019.145196 for further details.}
\end{figure}

\begin{figure}
	\centering
	\includegraphics[scale=0.6]{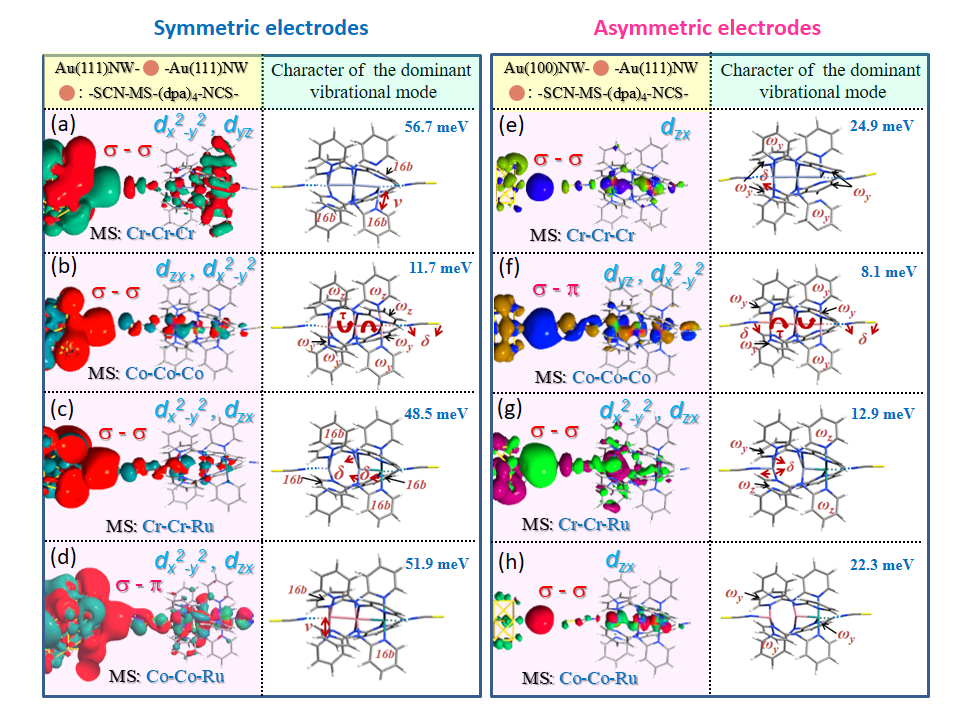}
	\caption{Please refer to https://doi.org/10.1016/j.apsusc.2019.145196 for further details.}       
\end{figure}

\begin{figure}
	\centering
	\includegraphics[scale=0.75]{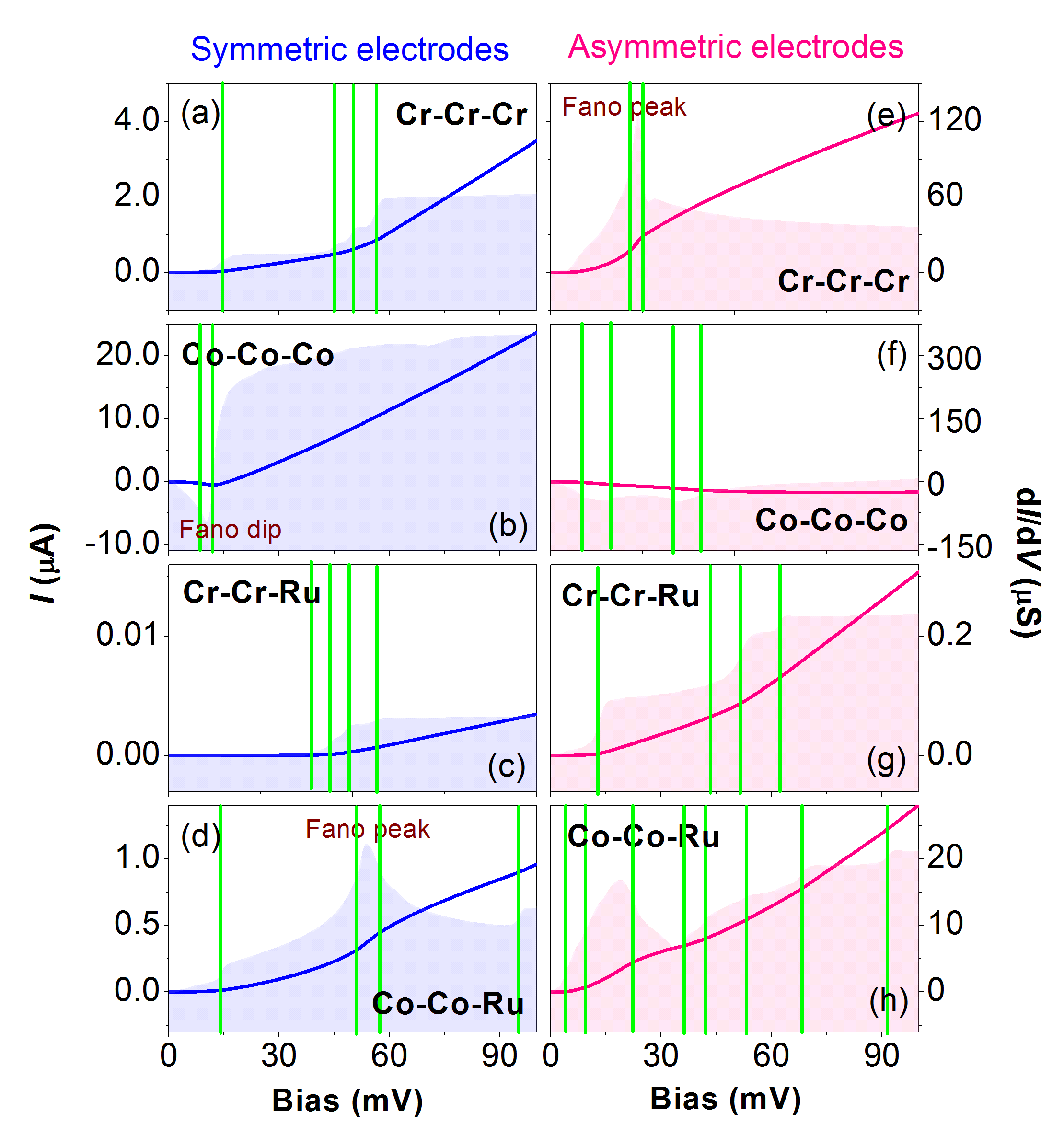}
	\caption{Please refer to https://doi.org/10.1016/j.apsusc.2019.145196 for further details.}  
\end{figure}

\begin{figure}
	\centering
	\includegraphics[scale=0.75]{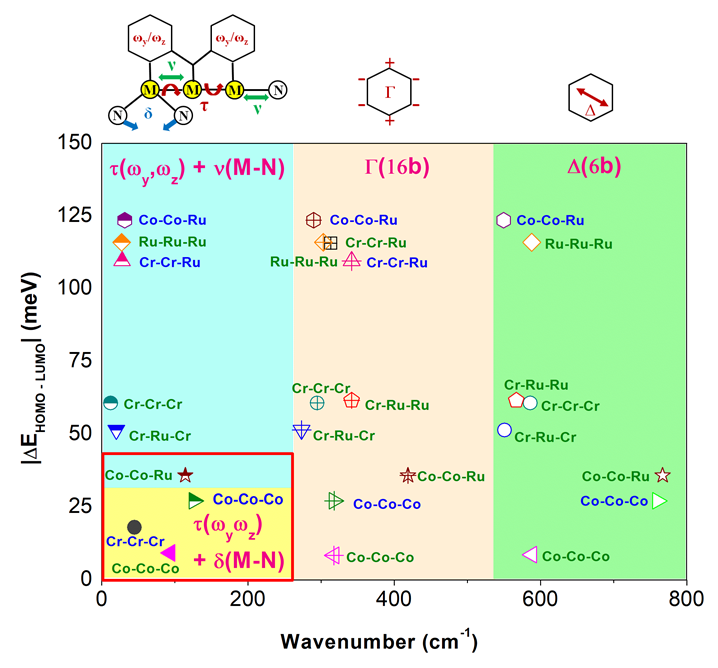}
	\caption{Please refer to https://doi.org/10.1016/j.apsusc.2019.145196 for further details.}  
\end{figure}

\begin{figure}
	\centering
	\includegraphics[scale=0.75]{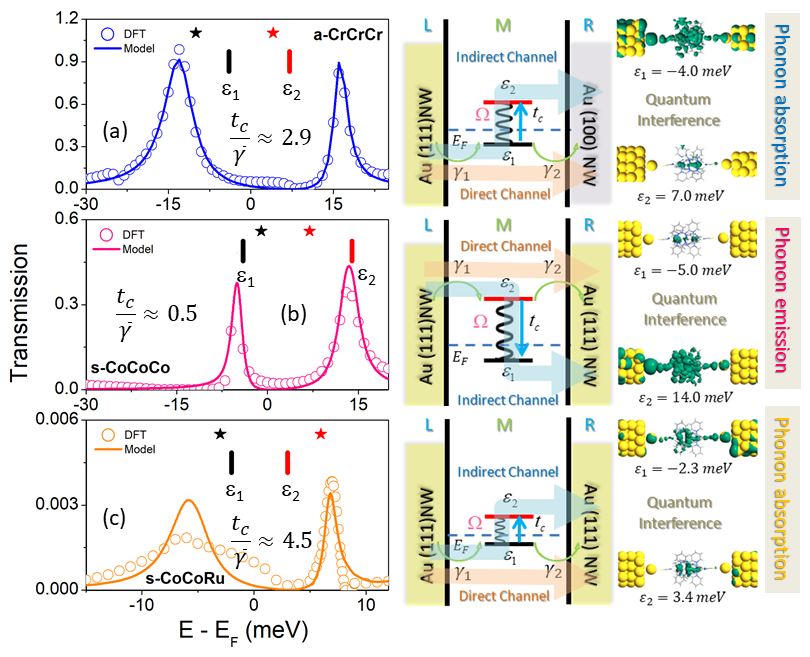}
	\caption{Please refer to https://doi.org/10.1016/j.apsusc.2019.145196 for further details.}  
\end{figure}

\begin{figure}
	\centering
	\includegraphics[scale=0.75]{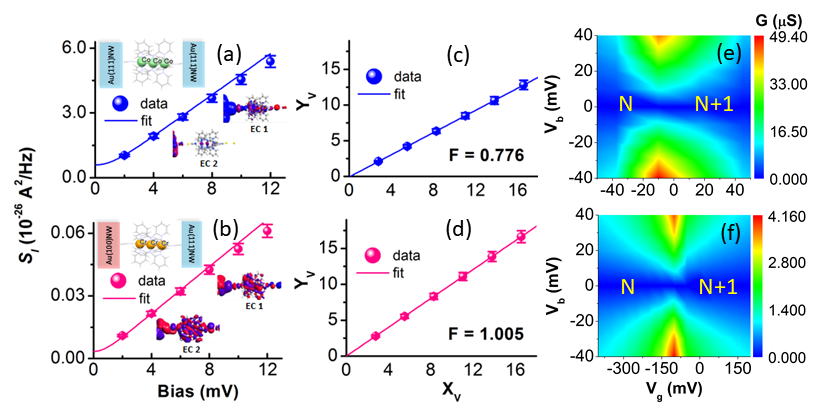}
	\caption{Please refer to https://doi.org/10.1016/j.apsusc.2019.145196 for further details.}  
\end{figure}

\begin{figure}
	\centering
	\includegraphics[scale=0.75]{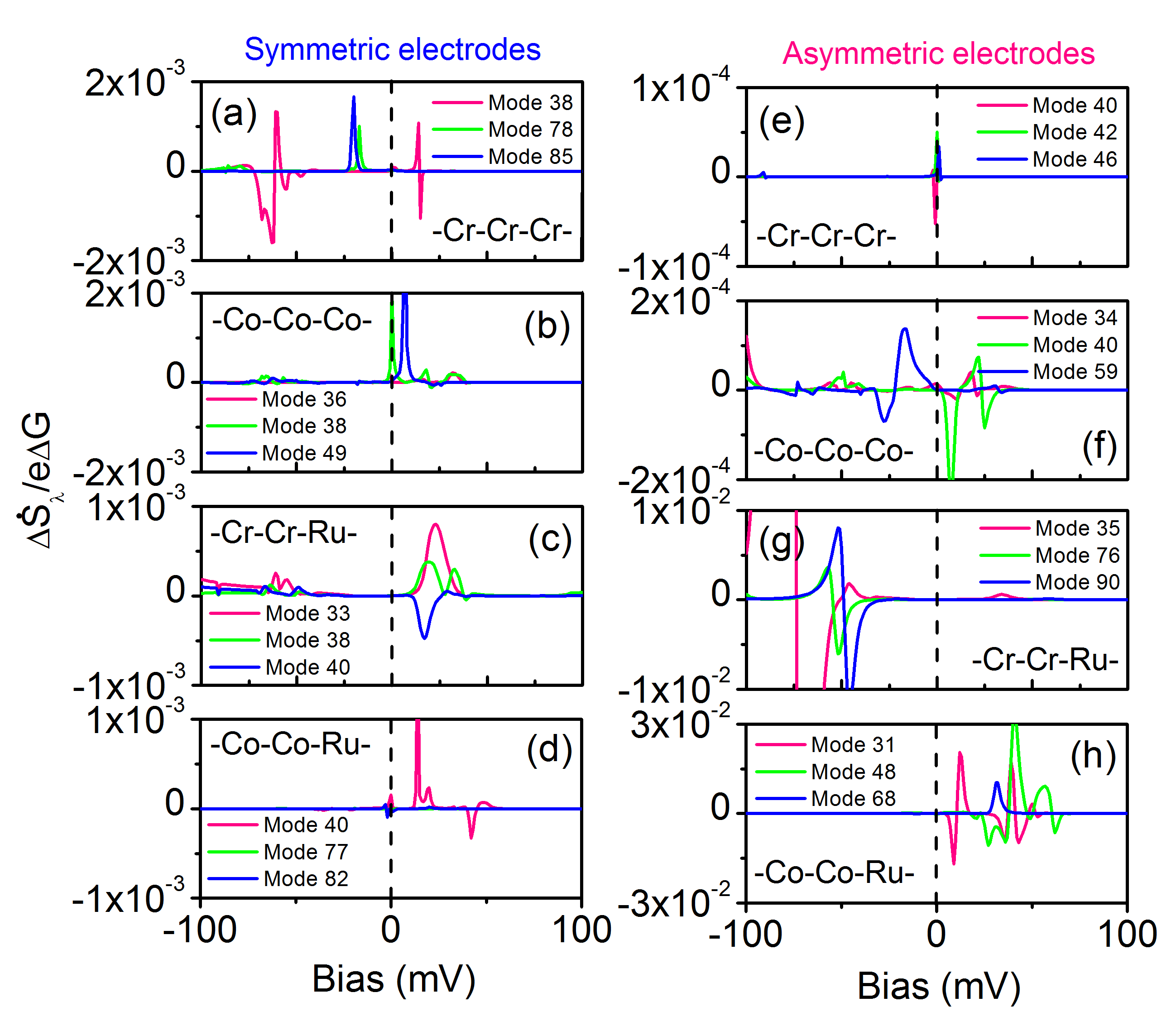}
	\caption{Please refer to https://doi.org/10.1016/j.apsusc.2019.145196 for further details.}  
\end{figure}

\section{Acknowledgement}
\scriptsize
This work was supported by the DST Nano Mission, Govt. of India, $via$ Project No. SR/NM/NS1062/2012. We acknowledge as well the DST-FIST, Government of India ($via$ Project No. SR/FST/PSI-155/2010) for providing the necessary computational resources. TRR further thanks the Council of Scientific and Industrial Research(CSIR), New Delhi, for her senior research fellowship (File no. 09/1045(0017)2K18). We are also thankful to SRM-HPCC, SRM Institute of Science and Technology for facilitating the high-performance computing.

\section*{References}

\bibliography{rebeda_arxiv_ic3}

\end{document}